# Exo-Jupiters and Saturns from two Gaia-like missions

By Erik Høg, Niels Bohr Institute, Copenhagen, Denmark

Erik.hoeg@get2net.dk

2013.09.06

**ABSTRACT:** Detection and orbit determination for thousands of planets with periods up to about 40 years would be obtained by astrometry from two Gaia-like missions, results which cannot be obtained by any other mission, planned or proposed. A billion stars of all spectral types will be surveyed. A comprehensive knowledge about heavy planets with these periods, reaching well beyond the snow line of any system, will lead to a better understanding of the formation and evolution of planets, also in the habitable zone. The theories of migration would obtain a better observational basis and we would learn whether the migration process is always active. Most other detections of exoplanets lie in the interval below 10 years, including those expected from the first Gaia. Astrometry with Gaia gives a large uniform sample of stars of all spectral types since it will not suffer from, e.g., the limitation of the radial-velocity method that only stars with sharp spectral lines can be measured. The two Gaia missions would therefore be complementary to the series of missions, realized or planned: CoRoT, TPF, WFIRST, Kepler, PLATO and NEAT which, with the words of ESA: "address one of the most timely and long-standing questions in science, namely the frequency of planets around other stars." – *Comments are welcome and further study is encouraged. The reader may think of other applications for the data from two Gaia missions, perhaps equally simple as this one; it took only a few days to describe the main points.*

1. Science goals for Gaia2

Two Gaia-like missions of 5 years duration launched at 20 year interval are considered in connection with the mission proposal by Høg (2013), a proposal submitted to ESA as supplement to the proposal by Brown (2013). But a 20 year interval should be assumed in performance estimates, not 15 years as in my mission proposal since ESA will not approve an astrometric mission so soon after Gaia.

Two missions with 20 year interval will give proper motions for a billion stars with 10 times smaller errors than from Gaia alone. This will change barely detectable kinematic phenomena on a few stars from, e.g., one Gaia mission into well-measured effects on samples of thousands of stars. I hope the reader would think of and describe such possibilities in order to set up *a list of science goals for a second Gaia-like mission*[1]. The linear proper motion is interesting, but the curved motion due to the gravitational force from a companion to the observed star is subject of the present report about discovery of exoplanets, but first a few words about double stars.

---

[1] After my recent visit to Strasbourg, R. Ibata wrote on 2 July: "…I would be very interested in contributing to this project, and will try to make a few simulations to get an idea just how much better one would do with Gaia2 for streams, dwarf galaxies and Local Group halo substructures..." –The report is here as Ibata (2013).

## 2. Double stars

Unresolved double stars can be discovered from astrometric observations in a single Gaia mission by the large residuals of the standard solution of linear motion for a single star. Many such discoveries were obtained with Hipparcos, in fact only 80 per cent of the stars could be resolved as single stars without problems (see Section 6.5.1 in ESA 2000). The corresponding expected percentage for Gaia is not known, according to information in June 2013 from Lindegren who wrote that it will probably not be too different from Hipparcos, i.e., around 80 per cent and less for bright stars.

The percentage for an interval of bright stars will be smaller because the astrometric errors are smaller and the stars are closer to us than the fainter stars. The longer Gaia mission of five years instead of three for Hipparcos will also make the percentage smaller.

A study by Söderhjelm (2005) is, according to Lindegren, the most recent covering the whole range of periods and detection methods, but it was based on simulations with a rather crude Galaxy model and old performance data for Gaia. Figure 5 in the paper gives the number of binaries detected and solved by different methods; comparing with the upper curve (the total number of binaries) it appears that between 5 and 20% of the binaries are detected, depending on the period. But as shown in Figs. 6 to 8 this varies a lot depending on magnitude and distance range.

With two missions a large fraction of stars can be discovered as binaries from the residuals of either mission and from a comparison of the proper motions from each mission and from the motions derived from the positions at the two epochs. Furthermore, the acceleration in the orbit can be determined. Such an analysis is illustrated in Figure 3. The additional use of Hipparcos results will lead to further discoveries, depending on the orbital periods.

## 3. Exoplanets from Gaia missions

A few exoplanets with periods longer than 15 years have been detected by radial velocities, timing of pulsars and direct imaging, but it will be shown here that thousands such planets with periods up to 40 years can be detected and measured with astrometry from two Gaia missions. Orbital periods of exoplanets about four times longer than from Gaia alone can be investigated. This should be expected from the ratio of 20/5=4 between the 20 year interval and the 5 year duration of each mission. The astrometric data from two Gaia missions would open a new parameter space of periods between 10 and 40 years in which other detection methods yield very little.

An important gain from two Gaia missions lies also in the sheer number of detected systems because the sample will be statistically very well defined, the selection criteria being given only by the astrometric characteristics of Gaia which are uniform over the entire sky. Gaia2 would be very efficient for deriving the frequency of heavy planets with periods in the Jupiter-Saturn range. This frequency is important for the understanding of the habitable zone because these heavy planets influence the formation and evolution of the inner parts of a planetary system.

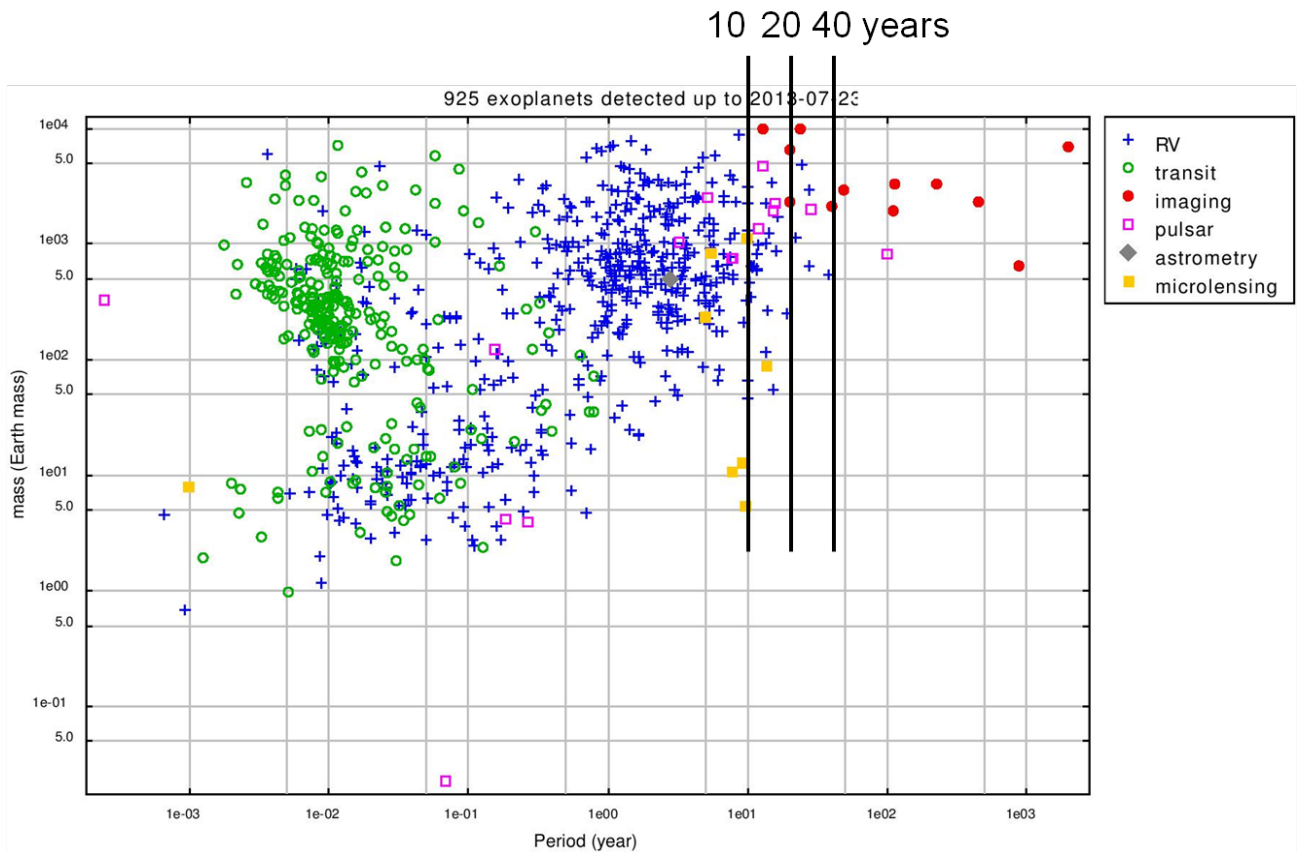

**Figure 1** Scatterplot showing masses and orbital periods of all extrasolar planets discovered through 2013-07-23, with colours indicating method of detection. There are 925 exoplanets and only one was detected by astrometry. The horizontal axis plots the log of the period, while the vertical axis plots the log of the mass. (Courtesy: Fabien Malbet, based on data from the site: "exoplanet.eu".)

The vast majority of known giant exoplanets orbit between their host star and the snow line. In current theories of planet formation heavy planets can only accumulate in the outer cold part of the system where water ice can exist. This means that they have migrated since their formation, going across and perturbing the habitable zone of the star. The present theories of migration have their observational basis only in the high frequency of giant exoplanets found close to the host star and in the observations of accretion disks around protostars. To find the frequency of planetary systems not affected by migration would be a important step in the estimation of the frequency of solar-like systems with habitable earth-like planets. The rate of "real Jupiter" and "real Saturn" will not provide new constraints to the migration models, but will indicate if these processes are usually very active (therefore, our Solar system would be an exception), or if they affect only a fraction of planetary systems.

In the solar system the snow line is about 5 AU, i.e. about an orbital period of 11 years. This implies that the first Gaia mission alone will not provide a large representative sample of exoplanets beyond the snow line, but also that the second Gaia will do so by reaching periods up to 40 years, see Figures 1 and 2.

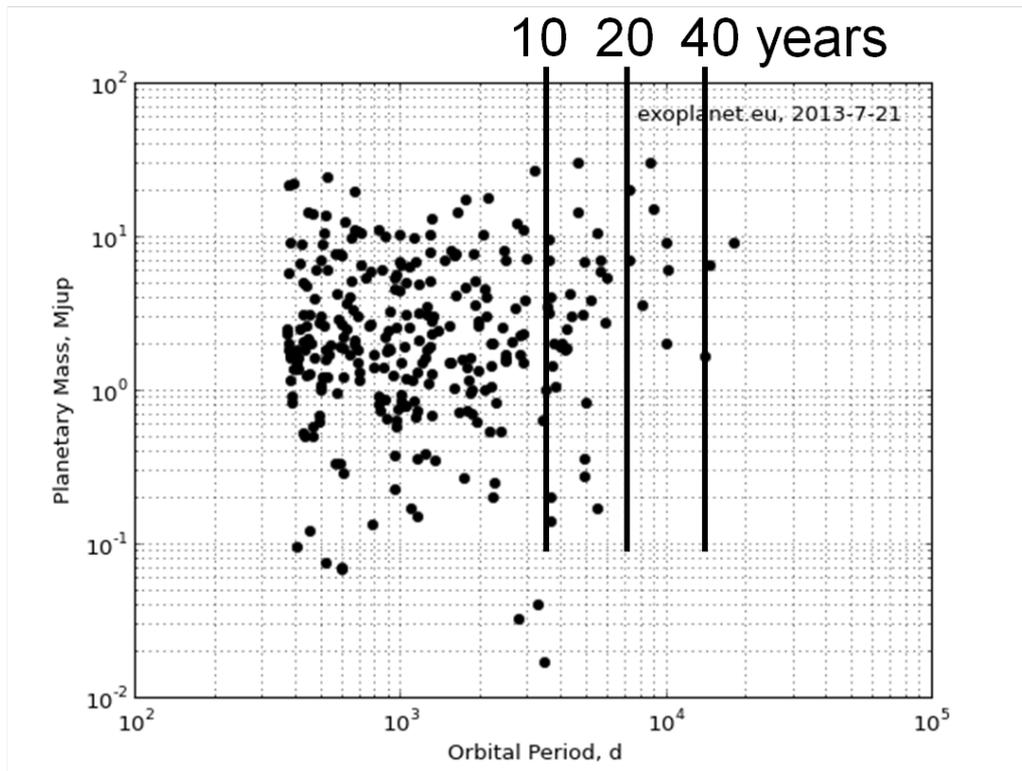

**Figure 2** The known exoplanets with periods between 1 and 100 years as of July 2013, re exoplanet.eu. The two Gaia missions will detect and measure thousands of exoplanets by astrometry with periods up to about 40 years, the first Gaia will cover up to about 10 years. The uniform coverage of the sky and of all spectral types will be a great advantage for our understanding of planetary systems.

Lattanzi et al. (2000) concluded from numerical experiments about Gaia1 that orbital parameters for over 2000 giant planets can be accurately estimated with the given assumptions with periods up to 11 years. It seems also quite plausible that periods twice the length of the mission can be detected and estimated. Applying the factor 4, we derive that periods up to about 40 years can be studied with two missions.

More recent information from Lattanzi (2009) says as follows. Within 200 pc of the Sun, and limiting counts to bright solar-type main-sequence stars, i.e. objects brighter than 13-th magnitude and with spectral types earlier than K5, about 300 000 objects are predicted to exist. With reasonable assumptions on the planetary frequency as a function of orbital radius, on the detection threshold, and on the accuracy of orbit determination, Gaia will be capable of discovering about 5000 giant planets around these stars. Gaia will accurately measure the orbital characteristics and actual masses better than 20 % for about 2000 of the detected systems.

Let us consider how many detected and measured planets would be required for a good characterization of the whole population in the period interval from 10 to 40 years. To characterize a planet the mass, the period, the orbit eccentricity, and the multiplicity of the system should be determined. Astrometry can give these four parameters, but not the diameter of the planet which is also desirable. In case of multiple systems, and we know they are frequent, three more parameters are needed to describe the orbits relative to each other: the orbit inclination, i, the longitude of the ascending node, Omega, and the argument of the pericentre, omega, see e.g. Perryman (2011). To characterize the host star we need especially the spectral type, the rotation, and the age, three parameters to be obtained with follow-up observations from the ground. The mass, metallicity, temperature etc. of the star should also be known, but should not be counted here as independent parameters. Altogether 4+3+3=10 parameters are desirable to characterize the population of planets. Some 1000-2000 planet detections would give a good picture of the population and this can be expected from Gaia1+2 data. Clearly, the capabilities of the present proposal should be quantified by a study with realistic simulations.

# Circular orbit  P= 30 yr, i=0

In a **simplified picture**, the residuals from Gaia1+2 contain 4 measurables from each mission: Orientation, length, curvature and acceleration
=> 8 measurables - if the accuracy is sufficient

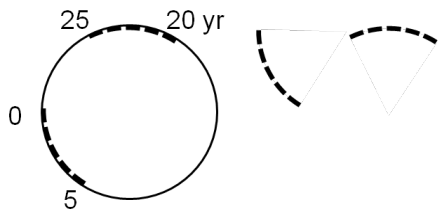

Specification of a *Keplerian orbit* requires 7 parameters: *a, e, P, t, i, Ω, ω*

Conclusion from a **deeper analysis**:
Gaia1+2 can provide *plenty* of data
Simulations are required to give an overview

**Figure 3** A circular orbit with period 30 years and inclination i=0 is shown in the upper part at left. To the right is shown two dotted curves, the same as on the orbit. Each dotted curve represents the residuals from each 5 year mission from a solution of the five astrometric parameters from all data. This simplified description gives 8 measured quantities as explained in the figure, if the accuracy is sufficient.

The orbit determination from Gaia1+2 observations is illustrated with the simplified description in Figure 3. Eight observed quantities are in principle obtained from the residuals in an astrometric solution for the two missions. This implies that the seven parameters of a Keplerian orbit projected on the sky could be determined – if the accuracy is sufficient and if, e.g., aliasing is not too strong. Clearly, simulations are required.

But this simplified description should be substituted by the following as planned for Gaia according to Halbwachs (2013-09-06, priv. comm.): "In an orbit calculation, the epoch is not counted among the measurable data, since it is used to derive the expected measurements and the partial derivatives of the orbital parameters.

An astrometric binary with a 30 year-period and a semi-major axis larger than around 10 mas will receive an acceleration solution from the CU4 processing. Therefore, Gaia1 will provide: the sum of the position of the barycenter + the mean orbital position for epoch 1, the parallax, the sum of the barycentric proper motion + the mean orbital velocity for epoch 1, the acceleration for epoch 1, and, possibly, the time derivative of the acceleration for epoch 1.

Gaia2 will provide again the parallax, and the same data as Gaia1, but for epoch 2. Except for the parallax, all the parameters are 2-D, and, for the two missions, the total is 13 "measurable data", or even 17 when the time derivatives of the acceleration are significant. An orbital solution contains 12 parameters: the 5 barycentric parameters + the 4 Thiele-Innes elements (or: a, i, Omega, omega) + P, e, T.

In brief, an orbital solution contains 12 parameters and the solution for a period of 30 years looks quite feasible with two Gaia missions since there are 13 measureable data in a real solution, and even 17 when the time derivatives of the acceleration are significant.

Detection of 10 mas orbits is of interest for binary stars, but not for finding exoplanets. In a combined solution of two missions, however, smaller orbits can be detected, and even smaller for shorter periods."

### 4. An overview of exoplanet detections

An overview of exoplanet detection is given by Horzempa (2012) to which the reader is referred for more details about the many space missions, realized or planned. A review on detection by astrometry is given by Malbet et al. (2009). Figure 1 shows the distribution of exoplanets as of 2013-07-23, a sample of the 925 exoplanets, belonging to 713 systems. 41 of the planets have periods longer than 10 years[2]: 12 of these were discovered by imaging, 1 by microlensing and 28 by RV. 17 of these 28 were member of systems with 2, 3 or 4 planets, i.e. the star had presumably been followed more intensely because one planet had already been found. The remaining 11 of the 28 planets were alone in their system. 34 of the 41 planets have periods less than 40 years.

Microlensing is a technique not suited for statistical investigations, since the properties of the host stars are not known. NASA is planning an infrared survey mission which will also be used to find exoplanets by microlensing, see NASA (2013a) and Horzempa (2012). The Wide-Field Infrared Survey Telescope (WFIRST) is a NASA observatory designed to perform wide-field imaging and slitless spectroscopic surveys of the NIR sky for the community.

Detection of exoplanets by transits has been very successful with the Kepler mission, but nearly all periods are shorter than 0.1 year according to the figure. The longest period is 303 days (Kepler-47c) according to NASA (2013), but this planet does not transit the stellar disk. The planned PLATO mission would detect periods by transits, especially of 1 year as that of the Earth. According to ESA in February 2010, PLATO would: "address one of the most timely and long-standing questions in science, namely the frequency of planets around other stars. This would include terrestrial planets in a star's habitable zone, so-called Earth-analogues. In addition, PLATO would probe stellar interiors by detecting the gaseous waves rippling their surfaces." ESA then foresaw a decision in mid-2011 and a launch no earlier than 2017. In October 2011, Solar Orbiter and Euclid were selected as the Cosmic Vision M1 and M2 missions while The Science Programme Committee decided to maintain the PLATO mission, not selected for a flight opportunity on this occasion, as a possible competitor for a future flight opportunity. PLATO is presented on ESA's website: ESA (2013).

The very few detections by radial velocity for long periods are summarized in the following section.

---

[2] Michel Mayor commented in an encouraging mail on 20 July: "Nice to read about your project for such a distant future space experiment. … Measurements during several decades done with modern spectrographs can easily detect the equivalent of Saturn. Obviously the advantage of Gaia2 will be the number of detections and the access to the mass."

## 5. Radial-velocity detection

The method is only suited for stars with many sharp lines, i.e. G and K dwarfs, until M2 in fact, with low rotation velocities. Radial-velocity observations of the star will only give a planet's minimum mass. If the planet's spectral lines can be distinguished from the star's spectral lines then the radial velocity of the planet itself can be found, and it is possible to derive the mass ratio of the system. Thus, the mass of the planet can be obtained when that of the star is derived from the spectral type.

The majority of planets in Figure 1 have been detected with the radial-velocity method, and most of these have periods below 10 years. The Extrasolar Planet Encyclopedia by Schneider (2013) contained in July 2013 a catalogue with 925 confirmed planets in 713 systems. The number of detected planets with periods between 10 and 40 years was 34 as given in Section 4, we shall now consider the upper part of this interval from 15 to 40 years. There were 17 planets in this interval, 10 of these had been detected by radial velocity between 2002 and 2013. The other 7 were detected by timing of pulsars (3) or by imaging (4), and none by microlensing, transiting or astrometry. 21 planets had been found with periods between 10 and 15 years. Of the 10 planets detected from radial velocities, 4 were alone in their system, while 6 belonged to systems with 2 or 3 planets.

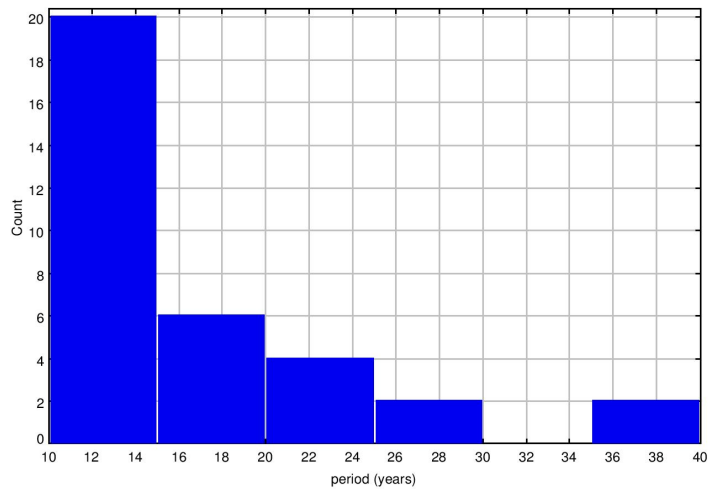

**Figure 4** The 34 detected exoplanets with periods from 10 to 40 years as of 2013-07-23. (Courtesy: Fabien Malbet, based on data from the site: "exoplanet.eu".)

All these numbers are given here to show how very few planets represent the interval from 10 to 40 years and Figure 4 illustrates the situation. Many more planets with long periods will be detected by radial velocity in the coming years, the present small number must be seen partly as a result of the short time, 18 years, since the first discovery was announced by Michael Mayor in 1996, and data reductions take time. The longest period is 14000 days found for 47 UMa d in a system around a V=5.1 mag star to which attention had been attracted by two other planets with periods of 1100 and 2400 days.

The host stars so far found by the RV method lie in the interval V=4 to 12 mag, see Figure 5. This may be compared with the 300,000 solar-type stars within 200pc and brighter than 13mag considered for Gaia in Section 3.

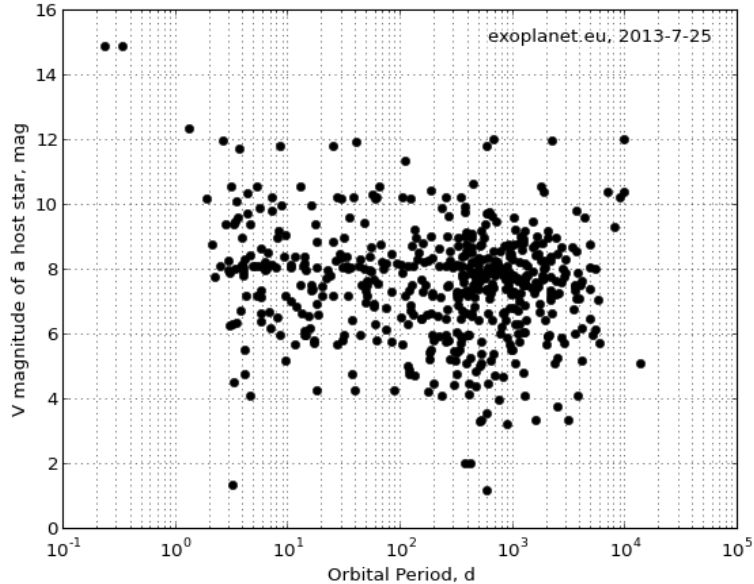

**Figure 5.** All exoplanets detected by radial velocity, plot of magnitude of host star vs orbital period.

A study should be made of the potential for collecting by radial velocities a representative and large sample of at least hundreds of planets with long periods. Does such a study perhaps exist already? This would be useful for a comparison with the astrometric detections with a Gaia2, taking into account that the Gaia2 astrometry will give thousands of detections,

An overview of the very high activity with the spectroscopic method is provided by the Tables 2.2 and 2.4 by Perryman (2011). The first table lists 34 spectroscopic instruments for past, present or future exoplanet searches. The second table lists 49 surveys focusing on particular stellar types. The number of target stars in a survey is listed, 7 of the surveys have 1000-2000 targets and one survey, the SDSS III MARVELS has 1000, but at the same time this survey uses an instrument with the smallest resolution of all, R=11000 while other present instruments have R=38000 to 115000.

A radial velocity survey comparable to the astrometry by Gaia1+2 should monitor 300,000 solar-type targets within 200pc during 20 years. But the Gaia1+2 astrometry always has the advantage over RV that it will provide the planetary mass without ambiguity, it will give the three orbit orientation parameters which are important for multiple systems, it will cover host stars of all rotations and spectral types, and it will serve other astrophysical purposes than detecting exoplanets, especially study the dynamics of stellar systems in our Milky Way and beyond. This leads to the conclusion that no other mission from ground or from space can cover the exoplanets with up to 40 year period as well as two Gaia missions.

Fabien Malbet has informed as follows: "The Kepler team has studied frequencies of planets, though not yet released the results, but one of the key results is that multiple systems are frequent. Therefore the best chance to find long period planet will be around stars which has already other planets detected. However one has to be cautious because the statistics are valid for short-period planets (less than 200d period), but if it is true, then the best probability to find long-period planet could be around stars that have already been identified as exoplanet host. This is one way of thinking, but the Gaia2 mission is indeed to check this argument with no a priori assumptions."

Later on, after reading the report, Fabien Malbet wrote the following: "Everything is fine and I find the report interesting, although I do think that for the next astrometry mission we should focus on a pointed astrometry mission to be able to follow up interesting Gaia candidates (exoplanets or others).” – And I agree, hoping that a technically feasible design of NEAT will soon be found, see Section 6.

## 6. On astrometric detection

The astrometric method is of course applicable to stars of all spectral types, an advantage over the Doppler-shift method. Historically, many claims were made for detection of companions by ground-based astrometry since the nineteenth century. A very prominent claim was made by an astronomer who excelled in the use of photographic astrometry to measure motions and parallaxes of many stars. Peter van de Kamp (1969) announced, as quoted in extenso from the abstract: "An alternate dynamical analysis of Barnard's star's motion over the interval 1938-1968 yields two companions in co-revolving, approximately coplanar, circular orbits with periods of 26 and 12 years, and masses of 1.1 and 0.8 times Jupiter, respectively." The claim was not confirmed by other observations and the full history of this case is told by Bell (2001).

The changes in stellar position are so small and atmospheric and systematic distortions so large that even the best ground-based telescopes cannot produce precise enough measurements. All claims of a planetary companion of less than 0.1 solar mass, as the mass of the planet, made before 1996 using this method are likely spurious.

Benedict et al. (2002) carried out astrometric observations of the system Gliese 876 using the Fine Guidance Sensor (FGS) instrument on the Hubble Space Telescope (HST) beginning around the time the first planet was announced and continuing for 2.5 years. Analysis of these data revealed a residual perturbation with semimajor axis of 0.3 ± 0.1 mas in phase with the orbit of planet b expected from modeling radial velocity data. This was the first definitive astrometric detection of an exoplanet.

One potential advantage of the astrometric method is that it is most sensitive to planets with large orbits. This makes it complementary to some other methods that are most sensitive to planets with small orbits. However, very long observation times will be required — years, and possibly decades, as planets far enough from their star to allow detection via astrometry also take a long time to complete an orbit.

Future space-based observatories such as Gaia will uncover new planets via astrometry, but for the time only one such planet has been confirmed. In 2010, Muterspaugh et al. published a study by ground-based astrometry of candidate substellar companions orbiting either the primary or secondary stars in several binaries. One of the star systems, HD 176051, a spectroscopic binary of 5.22 mag, was found to have a planet with "high confidence" level. The authors were however cautious and wrote: "Given that other astrometrically discovered substellar objects have not withstood the test of continued observations, these may represent either the first such companions detected, or the latest in the tragic history of this challenging approach." This paper thoroughly discusses all the advantages of the astrometric method over other methods. The discovery was confirmed, the orbital period of the planet, HD176051 b, is given as 1016 days in Schneider (2013), and Figure 1 shows the planet as the only one discovered by astrometry.

In the proposal by A. Quirrenbach for L2L3 (ESA 2013a) the following is found related to the NEAT mission: "An astrometric mission, which could for example be flown as ESA's M4 mission, could conduct an exhaustive search of the habitable zones of all nearby stars down to 1 $M_\oplus$, thus establishing the "ultimate" target sample for further exploration." The NEAT primary science program will encompass an astrometric survey of our 200 closest F-, G- and K-type stellar neighbors, with an average of 50 visits during a 5 year mission. NEAT is designed to carry out high-precision *differential* astrometric measurements at the 0.05 microarcsec (1 sigma) accuracy level in fields of 0.3 degree diameter. This is 100 times smaller error than Gaia is expected to obtain for *absolute* astrometry.

In August 2012, M. Shao, the original proposer of the NEAT concept, mentioned to me that ESA had concerns about the formation flying of the two spacecrafts required in NEAT and had advised the proposers to modify the design. The astrometric mission NEAT was proposed for M3, but the study report will be confidential as usual for responses to an Announcement of Opportunity and is therefore not provided on the ESA website. The concept is described in Malbet et al. (2012), NEAT (2013), and Malbet (2013). In the latter presentation the original concept is maintained as one option among other ones being considered, based on a review of all specifications.

## 7. Conclusion

It appears from Section 3 that two Gaia-like missions at 20 year interval will be well suited for detection of planets with all periods up to 40 years, around stars of all spectral types. The period interval of 10 to 40 years is very poorly covered by other methods of detection. During the past eleven years only nine planets have been detected by radial-velocity observations with periods between 15 and 40 years according to Schneider (2013) and seven were detected by timing or imaging. None of the nine space projects in the list found at Wikipedia, planned or proposed, nor any listed by Schneider (2013) will cover these long periods and it cannot be done from the ground with such uniformity and in such large number. Two Gaia-like missions will be unique since they will deliver a large and statistically well-defined sample of these planets.

I have met the objection against the proposal of a second Gaia that ESA is only interested in missions based on much new technology, but it is my experience from 32 years with ESA, 1975-2007, that the science carries very great weight. Imagine for a moment that the Schmidt telescope invented in the 1930s by Bernhard Schmidt had been used for some important astronomical purposes, but was then abandoned because it was not new anymore! When Gaia has demonstrated its capabilities the interest for a second mission will grow.

The popular scientific focus today is to search for possibly habitable planets, but equally important is to find how common, or seldom, such planets are in our part of the universe. This requires a deep understanding of the formation and evolution of planetary systems in general. For this purpose a large and representative sample of heavy exoplanets with long periods is needed because they have great influence on the evolution of the system. The two Gaia missions would therefore complete the series of missions, realized or planned: CoRoT – completed in 2013, Terrestrial Path Finder (TPF) – was a mission concept previously under study by NASA until 2012, WFIRST - planned by NASA (2013a), see also Horzempa (2012), Kepler – flying, PLATO - studied by ESA, and NEAT - studied by ESA which, with the words of ESA, "address one of the most timely and long-standing questions in science, namely the frequency of planets around other stars."

Planets figure in ESA's Cosmic Vision 2015-2025 Plan, which identified four scientific aims: "What are the conditions for life and planetary formation? How does the Solar System work? What are the fundamental laws of the Universe? How did the Universe begin and what is it made of?"

The combination of proper motions from Hipparcos, Gaia, and ground based astrometry could reveal a number of long period Jupiter-type massive exoplanets, but two Gaia missions will be orders of magnitude better because Gaia will obtain results with almost a hundred times smaller astrometric errors than Hipparcos.

**Acknowledgements:** I am grateful to Claus Fabricius, Jean-Louis Halbwachs, Peter Laursen, Lennart Lindegren, Morten Bo Madsen, Fabien Malbet, Michel Mayor, Aake Nordlund, Dimitri Pourbaix, Timo

Prusti, Andreas Quirrenbach, P. Kenneth Seidelmann, and Michael Shao for information and for comments to previous versions of this report, and I thank especially Jean-Louis Halbwachs and Fabien Malbet for very extensive and encouraging correspondence.